\documentclass[aps,prl,twocolumn]{revtex4-2}

\usepackage{amsmath}
\usepackage{commath}
\usepackage{graphicx}
\usepackage{float}
\usepackage{ulem} 

\usepackage[toc,page]{appendix}

\usepackage{graphicx}
\usepackage{dcolumn}
\usepackage{bm}

\usepackage{color}

\usepackage{CJKutf8}

\begin{document}


\title{Cold Dark Matter Based on an Analogy with Superconductivity}

\author{\begin{CJK*}{UTF8}{gbsn} Guanming Liang (梁冠铭)\end{CJK*}}\email{guanming.liang.25@dartmouth.edu}
\author{Robert R. Caldwell}\email{robert.r.caldwell@dartmouth.edu}

\affiliation{%
 Department of Physics and Astronomy, Dartmouth College, Hanover, New Hampshire 03755, USA}%
\date{\today}

\begin{abstract}

We present a novel candidate for cold dark matter consisting of condensed Cooper pairs in a theory of interacting fermions with broken chiral symmetry. Establishing the thermal history from the early radiation era to the present, the fermions are shown to behave like standard radiation at high temperatures, but then experience a critical era decaying faster than radiation, akin to freeze-out, which sets the relic abundance. Through a second-order phase transition, fermion-antifermion pairs condense and the system asymptotes toward zero temperature and pressure. By the present era, the nonrelativistic, massive condensate decays slightly faster than in the standard scenario--a unique prediction that may be tested by combined measurements of the cosmic microwave background and large scale structure. We also show that in the case of massive fermions, the phase transition is frustrated, and instead leaves a residual, long-lived source of dark energy. 

\end{abstract}

\maketitle


{\it Introduction}---The $\Lambda$CDM cosmological model is well supported by a wide range of observational and experimental data \cite{Planck:2018vyg,BOSS:2013rlg,DES:2021wwk,Brout:2022vxf,DESI:2024mwx}. Yet the nature of its two main ingredients, approximately $25\%$ cold dark matter (CDM) and $70\%$ cosmological constantlike dark energy ($\Lambda$), remains elusive to our understanding. 

The leading paradigm for dark matter posits a cold, nonrelativistic particle species that falls out of equilibrium with the primordial thermal bath as the Universe expands. The freeze-out mechanism and observed relic abundance of dark matter imply an electroweak energy scale for the interaction with the standard model (SM). However, laboratory searches at these scales for weakly interacting massive particles (WIMPs) have produced null results to date \cite{Roszkowski:2017nbc}. Nonthermal relics including axions \cite{Chadha-Day:2021szb} and primordial black holes \cite{Carr:2016drx} are actively sought through direct experimental detection \cite{ADMX:2018gho,Adams:2022pbo} and astrophysical observations \cite{DeLuca:2020agl,Bartolo:2018rku}, but remain at large \cite{Baer:2014eja,Carr:2020xqk}. The physics of dark matter is an open question.

The conventional explanation for the accelerated expansion of the Universe is a cosmological constant, $\Lambda$. However, naive estimates of the quantum vacuum energy are famously many orders of magnitude different from observation \cite{Weinberg:1988cp}, and $\Lambda$ is widely regarded as a placeholder until a deeper understanding of dark energy can be achieved \cite{Carroll:2000fy}. 

In this Letter, we propose novel candidates for CDM and dark energy as two different cases of a system of interacting fermions with broken chiral symmetry, similar to the condensation of Cooper pairs in the Bardeen-Cooper-Schrieffer (BCS) theory of superconductivity \cite{Bardeen:1957mv}.

The launching point of our investigation is the Nambu-Jona-Lasinio (NJL) model of interacting fermions \cite{Nambu:1961tp,Nambu:1961fr}. This theory was originally proposed as a relativistic model for dynamical mass generation, in analogy to nonrelativistic BCS superconductivity. More recently, the NJL model has been used in cosmological models of inflation \cite{Tong:2023krn}, dark matter \cite{Alexander:2016glq,Alexander:2024qml,Garani:2022quc,Alexander:2018fjp,Alexander:2020wpm}, dark energy \cite{Alexander:2006we,Alexander:2009uu,Inagaki:2003qq}, and more \cite{Tukhashvili:2023itb,Alexander:2008vt}. In most prior investigations, however, the model was either assumed to be at zero temperature or in equilibrium with the thermal bath of SM particles. That is, the distinct thermal behavior of NJL fermions was overlooked. Here, we emphasize the critical role of the thermal history for the proposed models of dark matter and dark energy. 

In brief, we present a scenario in which the relativistic NJL fermions decouple from the SM at early times, evolving like radiation with separate temperature from the thermal bath of SM particles. When the temperature drops past the axial chemical potential, the interacting fermion system begins to stiffen with an effective equation of state $w \equiv p/\rho = 1$, defined as the ratio of the homogeneous pressure to energy density. The duration of this stage sets the relic dark matter abundance, similar to the role of freeze-out for thermal dark matter. Below a critical temperature, the system undergoes a phase transition to the true vacuum, and the fermions coalesce into a cold, massive condensate as CDM. The standard big bang cosmology with dark matter proceeds as expected, from approximately GeV scales on down in the example we present.

In a second scenario, massive fermions are shown to spoil the phase transition, and the condensate is stuck in a long-lived, metastable vacuum as dark energy. The thermal history dictates that the equation of state  proceeds in stages as $w= 1/3\,\rightarrow\, 1\, \rightarrow\, -1$. In this view, the fine-tuning and coincidence problems are resolved in a way similar to models of quintessence \cite{Zlatev:1998tr}; the dark energy is set by the frozen dynamics of the condensate.

{\it Low Energy Theory}---The proposed theory of dark matter consists of fermion pairs $\bar\psi\psi$ each with Dirac mass $m$ and a quartic, scalar, attractive self-interaction with coupling $M>0$: 
\vspace{-1em}
\begin{align}\label{eqn:dirac}
\mathcal{L}&=\bar{\psi}(i\gamma^{\mu}\partial_{\mu}-m)\psi-\kappa\bar{\psi}\gamma^0\gamma^5\psi+\frac{(\bar{\psi}\psi)^2}{M^2}.
\end{align}
Since the Hubble expansion rate is negligible compared to the energy scales and parameters of the theory, we are justified to work in flat space, using temperature as our clock. Here $\kappa$ is the chemical potential associated with an axial asymmetry between left- and right-handed fermions: $n_L-n_R = \langle\bar\psi\gamma^0\gamma^5\psi\rangle \neq 0$. 
We note that the chirality flipping rate, $\Gamma_{\rm flip}\sim (m/3T)^2(T^5/M^4)$ \cite{Pavlovic:2016mxq,Boyarsky:2020cyk} is suppressed by the zero or small bare mass subsequently considered in this Letter. Hence, we may choose parameters such that $\Gamma_{\rm flip}$ falls below $H$ at energy scales higher than the validity of our model, and the axial charge is effectively constant. The $U(1)$ vector charge $\langle \bar\psi\gamma^0\psi \rangle=n-\bar n=0$ is neutral by construction. The Cooper pairs in this theory are fermion-antifermion pairs of like helicity, equal and opposite momenta, which are drawn together by the self-interaction. 

To proceed, we introduce the order parameter $\Delta$ and integrate out the fermions. Physically, $\Delta$ is interpreted as the energy gap. With the identity operator, $1=\int D\Delta\exp [ -\int d^4x (\frac{1}{2}{M\Delta}-\tfrac{1}{M}{\bar{\psi}\psi} )^2 ]$, we can perform a Hubbard-Stratonovich transformation on the fermion theory:
\vspace{-0.3cm}
\begin{equation}
    \mathcal{L}=-\frac{1}{4}M^2\Delta^2+\int\frac{d^4k}{(2\pi)^4}\ln\det\mathcal{G}^{-1}, \nonumber
\end{equation}
where the inverse Green's function is the Fourier space equation of motion $\mathcal{G}^{-1}=-\gamma^{\mu}k_{\mu}+\Delta-m-\kappa\gamma^0\gamma^5$. Redefining $\Delta-m\rightarrow\Delta$, the tree-level contribution with one-loop corrections is
\vspace{-.3em}
\begin{equation}
    \mathcal{L}= -\frac{1}{4}M^2(\Delta+m)^2+\int\frac{d^4k}{(2\pi)^4}\ln U_+ U_-, \nonumber
\end{equation}
where $U_{\pm} = \Delta^2-\omega^2+(k\pm\kappa)^2$. We evaluate the energy integral with Zeta regularization, but the momentum loop integral is divergent and we dimensionally regularize it with an ultraviolet cutoff $\Lambda_{UV}$. To leading order, we find the effective potential
\vspace{-.3em}
\begin{eqnarray}
    V_{\rm eff}(\Delta) &=& \frac{1}{4}M^2(\Delta+m)^2 + \frac{\Delta^4}{32 \pi^2}\left(1 + 4 \ln{\Delta}/{\Lambda_{UV}}\right) \cr 
    &-& \frac{\kappa^2\Delta^2}{4 \pi^2}\left(1 - 2 \ln {\Delta}/{\Lambda_{UV}}\right) + V_0.
    \label{eqn:vloop}
\end{eqnarray}
Self-consistency of this effective theory requires $M \gtrsim \Lambda_{UV} \gtrsim \kappa$. A striking feature of this model is the emergence of an exponentially suppressed energy scale. For sufficiently small mass $m$ 
such that $m \ll \Delta_0$,
there is an exponentially suppressed local minimum, the gap, at approximately 
\begin{equation} 
    \Delta_0 = \Lambda_{UV} \exp(-{\pi^2M^2}/{2\kappa^2}).
\end{equation} 
We lift the effective potential by $V_0$ so that the potential is zero at the minimum.

To understand the unique thermal evolution of the system, we calculate the finite temperature one-loop correction to the effective potential. The $\omega$ integral in the zero temperature theory is replaced with a Matsubara summation over discrete fermionic modes, $\omega_n=(2n+1) \pi/\beta$,  $d\omega/2\pi\rightarrow \frac{1}{\beta}\sum_{i\omega_n}$ and $\mathcal{G}^{-1}\rightarrow \beta\mathcal{G}^{-1}$. The one loop correction is
\vspace{-.3em}
\begin{equation}
    \mathcal{L}_1(\Delta, T) = \int\frac{d^3k}{(2\pi)^3}\frac{1}{\beta} \sum_{n=-\infty}^\infty  \ln\beta^4 U_{n+}U_{n-} \nonumber
\end{equation}
where $U_{n\pm} = \Delta^2 + \omega_n^2+(k\pm\kappa)^2$.
With the complex identity $\lim_{\delta\to 0}\sum_n {\rm e}^{i\omega_n \delta} \ln(\beta s - i \omega_n \beta) = \ln(1 + {\rm e}^{-\beta s})$, the theory can be separated into a zero temperature contribution as well as finite temperature corrections:
\vspace{-.2em}
\begin{equation}
     V_{\rm eff}(\Delta,\,T) = V_{\rm eff}(T=0) -\frac{2}{\pi^2\beta^4} [ I_2(\beta\Delta) + \kappa^2 \beta^2 I_0(\beta\Delta) ] \label{eqn:vt}
\end{equation}
where $I_n(x)=\int_{0}^{\infty}ds\, s^n\ln [ 1+e^{-\sqrt{s^2+x^2}} ]$.

\begin{figure}[h]
    \centering
    \includegraphics[width=1.0\linewidth]{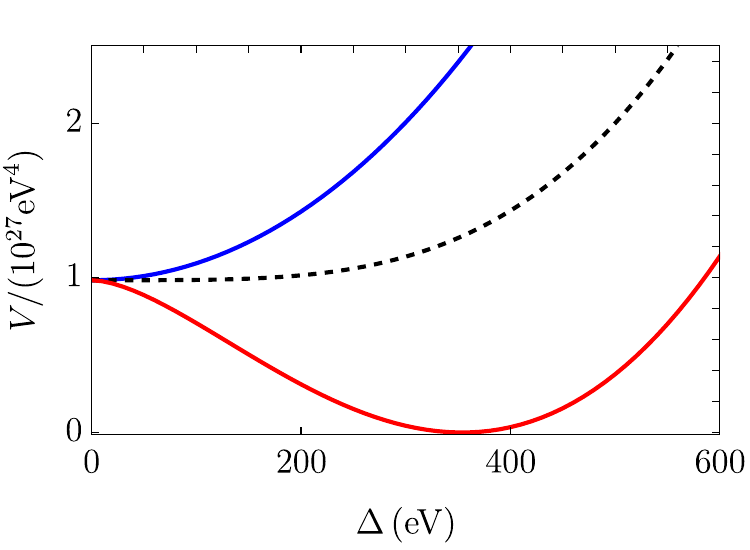}
    \caption{Effective potential $V(\Delta,T)$ versus $\Delta$ for the massless theory at $T=2 T_c$ (blue), $T=T_c$ (black, dashed), and $T=0$ (red). The parameters are $(\kappa,\,\Lambda_{UV},\,M)=(0.56, 1.1, 1.2)$~TeV, for the cold dark matter scenario. For illustrative purposes, the $T\neq 0$ curves have been artificially shifted vertically to match the $T=0$ curve at $\Delta=0$.}
    \label{fig:Veff}
    \vspace{-0.2cm}
\end{figure}

Inspection of Eqs.~(\ref{eqn:vloop})--(\ref{eqn:vt}) reveals that the gap field for the massless theory undergoes a phase transition at temperature $T_c$, as illustrated in Fig.~\ref{fig:Veff}. The boundary condition for the critical temperature is $\Delta(T_c)=0$, with gap equation given by $\partial V_{\rm eff}/\partial\Delta = 0$. It is straightforward to obtain $T_c = \Delta_0 e^{\gamma_e} /\pi$ where $\gamma_e$ is the Euler-Mascheroni constant; the factor $e^{\gamma_e}/\pi$ is reminiscent of s-wave superconductors in BCS theory.

Above a critical temperature, $T> T_c$, the gap field $\Delta$ sits at zero. For $T< T_c$ in the massless case, $m=0$, the field undergoes a second-order phase transition, where the minimum smoothly transfers from $\Delta=0$ to $\Delta_0$. We will consider this case for the dark matter scenario.

We note that in BCS theory only fermions near the Fermi surface feel the attractive interaction. Whereas in this relativistic NJL theory, all fermions with momenta below the cutoff $\Lambda_{UV}$ feel the interaction. As a result, the fraction of fermions that are in pairs is $\Delta/\Delta_0$, which rapidly approaches unity as $T$ drops below $T_c$.

In the massive case with 
$m\lesssim\Delta_0$, 
the gap field is stuck at $\Delta=0$, trapped behind a barrier, metastable to a true minimum near $\Delta_0$. We consider this case for the dark energy scenario.

Note that we have ignored the possibility of a quartic pseudoscalar interaction, $(\bar\psi i \gamma^5 \psi)^2$ in Eq.~(\ref{eqn:dirac}). For this case, the effective potential is the same as for the scalar four-fermion case, except the factor $(\Delta+m)^2$ is replaced by $\Delta^2 - m^2$ in Eq.~(\ref{eqn:vloop}). This $m^2$ term gets absorbed into $V_0$, and as a result, the pseudoscalar case is identical to the massless scalar case.

{\it Thermal History}---Now we can begin to sketch the thermal history. We follow the standard definitions of pressure and energy density, $p=-V$ and $\rho=Tdp/dT-p$ \cite{Kolb:1981hk}. At high temperatures, $T \gg \kappa, m$, the gap field lies at the potential minimum, $\Delta = 0$. The quantity $I_2(0)$ evaluates to $7 \pi^4/360$, which yields $p(T, \Delta=0)=\frac{7}{8}\frac{2\pi^2}{45}T^4=\frac{1}{3}\rho$. This agrees with the expected energy density and pressure of a relativistic gas of free fermions, accounting for both particle and antiparticle, and two helicities.

We assume these fermions were once in thermal equilibrium with the SM, sharing the same temperature $T_{\Delta}=T_{\gamma}$ as photons. We further assume scattering processes maintaining equilibrium later became inefficient and the fermion system decoupled, so that $T_\Delta$ evolved separately, adiabatically thereafter: $s_\Delta=(\rho_\Delta+p_\Delta)/T_\Delta \propto a^{-3}$, where $a$ is the cosmic expansion scale factor. The subsequent dropout of SM species from equilibrium heats the photons so that $T_\gamma > T_\Delta$. The presence of this additional, relic radiation contributes to the rate of cosmic expansion, which in turn is tightly constrained through observations of the light element abundances predicted by big bang nucleosynthesis (BBN), and through observations of the cosmic microwave background (CMB). These constraints can be satisfied by enforcing a sufficiently early time for decoupling. For example, decoupling prior to the QCD phase transition would be adequate to lower $T_\Delta$ relative to $T_\gamma$ to satisfy the Planck 2018 bound $\Delta N_{\rm eff} < 0.3$ (95\% CL, one-tailed Planck TT,TE,EE+lowE+lensing+BAO \cite{Planck:2018vyg}) on the effective number of additional, relativistic particle species, as customarily measured in units of massless neutrinos.
 
As the temperature drops below $\kappa$, and the gap remains at $\Delta = 0$, we see that the effective potential is dominated by the $I_0$ term in Eq.~(\ref{eqn:vt}), where $I_0(0)=\pi^2/12$. The resulting energy density and pressure are $p_\Delta = \rho_\Delta = \kappa^2 T^2/6$ with $w_\Delta=1$. For this equation of state, adiabatic evolution yields $\rho_\Delta \propto (1+z)^6$, where $z$ is the expansion redshift. Hence, the $\Delta$ field decays faster than both radiation and nonrelativistic matter. This $\kappa$-dominated evolution plays a comparable role to freeze-out for thermal WIMPs, dropping the field out of equipartition with the cosmic fluid, and setting the relic density for the subsequent matter-dominated evolution.

\begin{figure}[h]
    \centering
    \includegraphics[width=1.0\linewidth]{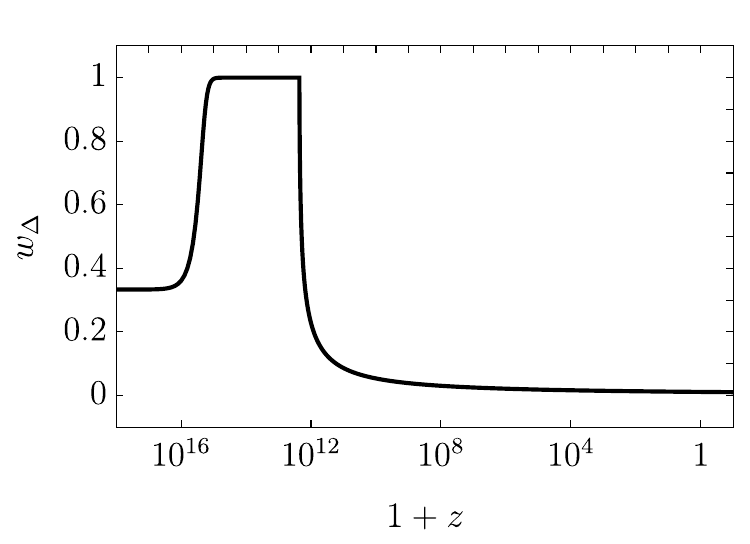}
    \caption{Equation of state $w_\Delta$ versus redshift, $z$. There are four distinct eras: Radiation, $w_\Delta=1/3$; $\kappa$-domination, $w_\Delta = 1$; Phase Transition, at $z\sim 10^{12}$; Pressureless matter, at $z\lesssim 10^{12}$. Same parameters are used as in Fig.~\ref{fig:Veff}.}
    \label{fig:w}
\end{figure}

\begin{figure}[h]
    \centering
    \includegraphics[width=1\linewidth]{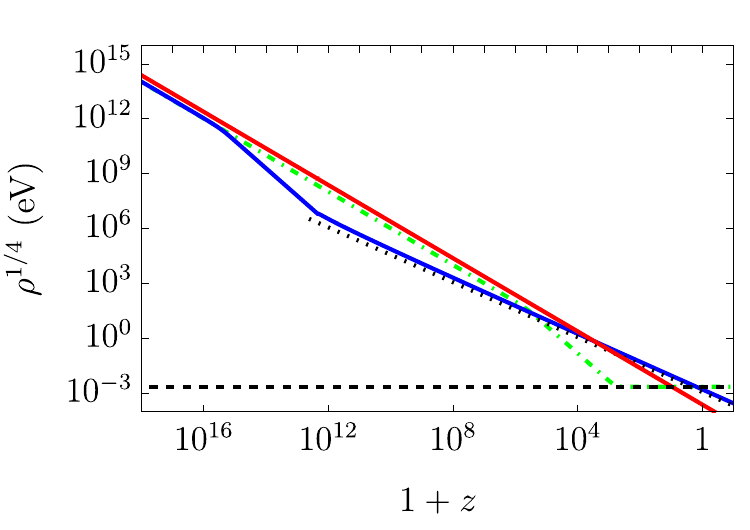}
    \caption{Cosmic energy density $\rho^{1/4}$ vs. redshift, $z$. Standard Model radiation (red), $\Delta$ field with $\Omega_{\rm CDM}=0.25$ (blue), cosmological constant dark energy with $\Omega_{\rm DE}=0.7$ (black dashed), and baryons with $\Omega_{\rm B}=0.05$ (black dotted). Same parameters for CDM are used as in Fig.~\ref{fig:Veff}. The proposed, metastable dark energy model is also shown (green, dot-dashed).}
    \label{fig:rho}
    \vspace{-0.2cm}
\end{figure}

The $\kappa$ phase ends when the temperature reaches $T_c$. For the massless case, the field undergoes a second-order phase transition, where the minimum smoothly transfers from $\Delta=0$ to $\Delta_0$ as $T \to 0$. To model this behavior, the integrals in Eq.~(\ref{eqn:vt}) must be evaluated numerically. However, for $T \lesssim T_c/2$, we observe $w_\Delta \simeq T/\Delta_0$. As $T$ drops well below the gap field strength, the condensate approaches the state of ideal, pressureless matter. This suggests an analytic solution, 
\begin{equation}
    (1+z)^3 = \frac{w_0 (1+w_\Delta)}{w_\Delta(1+w_0)}\, e^{\tfrac{1}{w_0}-\tfrac{1}{w_\Delta}} ,
    \label{eqn:w}
\end{equation}
where $w_0$ is the present-day value of $w_\Delta$. For the example model, $w_0\simeq 0.01$ and $w_\Delta \lesssim 0.02$ at equality. Recent studies of the dark-matter equation of state \cite{Kopp:2018zxp,Ilic:2020onu} based on current CMB and large scale structure (LSS) data such as Planck \cite{Planck:2018vyg} and the Sloan Digital Sky Survey \cite{BOSS:2013rlg} find a subpercent level constraint on $w$ near recombination at the $99\%$ confidence level. However, we caution that the studies assumed $w$ in bins joined by sharp transitions, in contrast with the slow, continuous evolution predicted in Eq.~(\ref{eqn:w}). The equation-of-state history and energy density in our model are illustrated in Figs.~(\ref{fig:w}-\ref{fig:rho}).

{\it Coldness of Dark Matter}---To this point, we have employed a mean field approximation to characterize the statistical average behavior of the fermionic ensemble. In the approach to $\Delta_0$, we must consider fluctuations about the minimum. Here, the group velocity of these fluctuations, on scales relevant for the LSS, characterizes the coldness of the dark matter \cite{Armendariz-Picon:2013jej}. 

Perturbing the gap field, $\Delta \to \Delta_0 + \delta\Delta$, we consider fluctuations of the potential energy
\begin{equation}
    V(\Delta)\simeq V(\Delta_0)+\tfrac{1}{2}{V''(\Delta_0)}(\Delta-\Delta_0)^2= \tfrac{1}{2}(\kappa \delta\Delta/\pi)^2
\end{equation}
as well as kinetic terms $\xi^{\mu\nu} \partial_\mu \delta\Delta \partial_\nu \delta\Delta$, following similar calculations in Ref.~\cite{Tong:2023krn}. The coefficients $\xi^{\mu\nu}$ are given by the $k^2$ derivatives of the fermion self-energy amplitude at finite temperature: 
\vspace{-.4em}
\begin{eqnarray}
    \mathcal{M}&=&-\frac{1}{\beta}\sum_{i\omega_n}\int\frac{d^3p}{(2\pi)^3}\text{Tr}\left[i \mathcal{G}(p)i \mathcal{G}(p+k)\right]\cr
    \xi^{00}&=&\left.\frac{d\mathcal{M}}{dk^2_0}\right\vert_{k=0}\text{and}\hspace{.08in}\xi^{33}=\left.\frac{d\mathcal{M}}{d\abs{k}^2}\right\vert_{k=0} \label{eqn:xi}
\end{eqnarray}
where $\mathcal{G}^{-1}(p)$ is as given previously. In the $T\to 0$ limit, $\xi^{00} = -3\xi^{33} = \kappa^2/12 \pi^2\Delta^2$, to leading order. In Fourier space, the dispersion relation is $\omega^2 = \tfrac{1}{3} k^2 + 12 a^2 \Delta_0^2$. The physically relevant group velocity is $v_g = d\omega/dk \sim {\cal O}(k/a \Delta_0)$, which is negligible on cosmological scales. 
 
At finite temperature, the momentum integral in ${\cal M}$ must be evaluated numerically. Using the same TeV-scale parameters as in Figs. \ref{fig:Veff}-\ref{fig:rho}, the finite temperature corrections to $\xi^{00},\, \xi^{33}$ are found to be exponentially suppressed. For example, the thermal corrections to $\xi^{00}$ are well-described by $|1 - \xi^{00}_T/\xi^{00}| \simeq (x \ln x)^2 e^{-x}$ for $x = \beta\Delta$ in the range $x=5 - 30$; for larger values of $\beta\Delta$ the exponential suppression dominates. This run of $\beta\Delta$ covers the epoch $z \lesssim 10^{12}$, relevant for the CMB and LSS. 
For the example model, the excitations are relativistic until $z \sim 10^{12}$, after which $\Delta_0 \sim 100$~eV. The comoving free-streaming length $r_{FS} = \int v_g dt/a$ is therefore a tiny $10^{-7}$~Mpc, which is well below the scales probed by observations of the CMB and Ly-$\alpha$ forest \cite{Armendariz-Picon:2013jej}.

In this initial foray, we have not attempted to embed our low energy model in a realistic particle theory. However the axial asymmetry, necessary for the existence of the nontrivial gap solution, suggests a possible connection to leptogenesis models that explain the matter-antimatter asymmetry of the Universe \cite{Davidson:2008bu}. Around $\Delta_0$, the thermal ensemble average of the axial number density $n_A=\langle \bar\psi \gamma^0 \gamma^5 \psi \rangle$ and the overall fermion number density $n_{\Delta}=\langle \bar\psi \psi \rangle$ are given by:
\begin{eqnarray}
    n_A&=&-\frac{\partial V(\Delta_0,T)}{\partial \kappa}=\frac{\kappa\Delta_0^2}{2\pi^2}\left(1+\frac{\pi^2M^2}{\kappa^2}\right)+\frac{\kappa T^2}{3} \cr
    n_{\Delta} &=&\left.\frac{\partial V(\Delta,T)}{\partial \Delta}\right\vert_{\Delta_0}=\frac{1}{2}M^2\Delta_0.
\end{eqnarray}
To quantify the degree of asymmetry, we define a chiral asymmetry parameter $r_A \equiv n_A / n_\Delta$. By the present day, with $T \ll \Delta_0$, $r_A=\frac{\Delta_0}{\kappa}\left(1+{\kappa^2}/{\pi^2 M^2}\right)$. The exponential suppression of $\Delta_0$ easily facilitates parameters such that 
$r_A \simeq \eta$, where $\eta \equiv n_B/n_\gamma = 6.10(\pm 0.04) \times 10^{-10}$ \cite{Planck:2015fie} is the baryon asymmetry parameter.
In principle, given a leptogenesis scheme that connects $r_A$ to $\eta$, then if $M$ and $\kappa$ are fixed, one can use the observed baryon-to-photon ratio to determine the cutoff, $\Lambda_{UV}$.

{\it Dark Energy}---We consider the massive NJL fermion condensate as a candidate for cosmological constantlike dark energy. We restrict attention to the case $m\lesssim\Delta_0$, whereby the thermal history of massive NJL fermions is similar to the massless case during the periods $w_\Delta=1/3$ and $w_\Delta=1$. However, the $\tfrac{1}{4}M^2(\Delta+m)^2$ term in Eq.~(\ref{eqn:vloop}) has positive slope at the origin, resulting in a potential barrier that traps the field at $\Delta=0$ at low temperatures. 

As before, we shift the potential such that $V(\Delta_m)=0$, where $\Delta_m\simeq\Delta_0 + m \ln(\Delta_0/\Lambda_{UV})$ is the global minimum. Instead of undergoing a phase transition at $T_c$, the field remains in a potential-dominated metastable state with equation of state $w_\Delta \to -1$. For an example model, we use $(M,\,\Lambda_{UV},\,\kappa,\,m) \sim$ $(128,\, 96,\, 80,\, 7.5\times 10^{-6})$~eV to obtain $\Omega_{\rm DE} \simeq 0.7$ as shown in Fig.~\ref{fig:rho}. This metastable state is separated from the true vacuum by a barrier $\sim 10^{-6}$~eV${}^4$ high and $\sim 10^{-4}$~eV wide. To estimate the lifetime of this state, we consider the bubble nucleation rate $\Gamma = A\exp(-S_E)$, where $S_E$ is the Euclidean action of the tunneling path joining the metastable and true vacuums. For the chosen parameters, we are justified to work in the thin-wall limit, and $S_E$ may be approximated \cite{Kolb:1981hk}
\vspace{-.4cm}
\begin{equation}
S_E=\frac{27\pi^2\left[\int_{0}^{\Delta_m}d\Delta\sqrt{2 V(\Delta)}\right]^4}{2\left(V(0)-V(\Delta_m)\right)^3}\simeq 2.8\times 10^7.
\end{equation}
For any reasonable value of $A$, a first-order phase transition to $\Delta_m$ is exponentially suppressed and the time to tunnel is much longer than the age of the Universe.

For this model, the $\kappa$ phase begins when the SM photon temperature reaches $T_\gamma \sim 1$~keV, near $z\sim 10^6$. Decoupling of the fermions from the SM must occur sufficiently early to satisfy BBN constraints on the excess degrees of freedom of the cosmic fluid. The $\kappa$ phase ends and the field is potential dominated from $z\sim 10^3$ onward. This model also addresses various problems that plague many dark energy models \cite{Velten:2014nra}. At early times the fermions are in thermal equilibrium with the SM, so no fine-tuning of initial conditions is required. The exponential suppression of $\Delta_m$, and the requirement that $m \ll \Delta_m$, necessary for the existence of a minimum, helps explain the smallness of dark energy. And the chemical potential, which separates CDM and dark energy from the thermal bath, helps explain the coincidence of energy densities.

{\it Conclusions}---We present a novel candidate for CDM and dark energy. In analogy with superconductivity, the dark matter particles are nonrelativistic Cooper pairs of massless fermions. For dark energy, the potential energy is associated with pairs of massive fermions. We highlight the important role of the thermal history, in particular the mechanism of the chemical potential, in setting the relic abundance. The scenario further benefits from the exponentially suppressed physical scales, well below the cutoff of this low energy theory, that are generated by the gap field solution. The exponential suppression may help explain two small numbers in cosmology: the smallness of dark energy, and the axial asymmetry which may link dark matter to the baryon asymmetry. A unique prediction of the model is a nonzero, time-evolving equation of state, $w_\Delta$. CMB and LSS data may be able to constrain $w_\Delta$ down to the percent level across certain redshift bands \cite{Kopp:2018zxp,Ilic:2020onu}, and thereby test the energy scales and interactions of the fermion system. More precise constraints on $\Delta N_{\rm eff}$ expected from future CMB missions such as the Simons Observatory \cite{SimonsObservatory:2018koc} and CMB-S4 \cite{CMB-S4:2016ple} will further inform the energy scales and decoupling of these species. The slightly faster decay of CDM will also shift the geometry of the CMB acoustic oscillations, and may help relieve another outstanding problem of cosmology, the Hubble tension \cite{Freedman:2017yms}. We reserve for future work to investigate the imprint on the CMB and LSS, to consider the collisional nature of this species of dark matter, and embed this low energy theory in a more complete model of particle physics.

{\it Acknowledgments}---We thank Rufus Boyack, Yuhang Zhu, and Yi Wang for helpful discussions and insight. G.L. was supported by the Presidential Scholarship and Wilder Fellowship at Dartmouth College.

\bibliography{main}

\end{document}